\documentclass{PoS}

\title{Excited nucleon spectrum with two flavors of dynamical fermions}

\ShortTitle{$N_f=2$ excited nucleon spectrum }
\author{John M. Bulava$^a$,  Robert G. Edwards$^b$, \speaker{Eric Engelson}$^c$,
          B\'alint Jo\'o$^b$,  Adam Lichtl$^d$, Huey-Wen Lin$^b$,
        Nilmani Mathur$^e$, Colin Morningstar$^a$,  David G. Richards$^b$, and
         Stephen J. Wallace$^c$}
     \author{for the Hadron Spectrum Collaboration\\
        \llap{$^a$}Department of Physics, Carnegie Mellon University, Pittsburgh, PA 15213\\
        \llap{$^b$}Thomas Jefferson National Accelerator Facility, Newport News, VA 23606\\
        \llap{$^c$}Department of Physics, University of Maryland, College Park, MD 20742\\
        \llap{$^d$}RIKEN-BNL Research Center, Brookhaven National Laboratory, Upton, NY 11973\\
        \llap{$^e$}Department of Theoretical Physics, Tata Institute of Fundamental Research, India\\
        }

\abstract{We compute the spectrum of excited nucleons  using the anisotropic Wilson lattice with
two flavors of dynamical fermions.  Using optimized sets of
operators which transform irreducibly under the octahedral group,
 matrices of correlation functions are computed.  We apply the
variational method to these matrices to extract excited energy
eigenstates.  We obtain several states for each irrep and identify
the continuum spin for the lowest-lying states, including a
$J^P=\frac{5}{2}^-$ state.}

\FullConference{The XXVI International Symposium on Lattice Field Theory \\
                July 14 - 19, 2008\\
                Williamsburg, Virginia, USA}

\begin{document}

\section{Introduction}
The study of the excited baryon spectrum is an important research
program for lattice QCD.  In this report we present preliminary
findings for the nucleon excited spectrum using the $N_f=2$ anisotropic Wilson action.  We see
evidence for a $J^P=\frac{5}{2}^-$ state.

We construct operators which transform according to the
irreducible representations (irreps) of the lattice rotation group
(the octahedral group for cubic lattices)
\cite{basak074,basak094}. There are six double-valued irreps:
$G_{1g}$, $H_g$, $G_{2g}$, $G_{1u}$, $H_u$, and $G_{2u}$, where the
``g'' subscript denotes positive parity (gerade) irreps and the
``u'' subscript denotes negative parity (ungerade) irreps. We
identify the continuum limit spins by subducing the continuum
rotation group to the octahedral group. The pattern of continuum
spin states in each octahedral irrep is summarized in Table
\ref{tab:irreps}.

To extract the excited spectrum, we construct a large number of
operators in each symmetry channel.  It is necessary to include
operators with quarks displaced with respect to one another to
capture radial and orbital excitations.  After
selecting a manageable set of optimized operators, we use the
variational method, diagonalizing the matrix of correlation
functions, to extract the excited spectrum.  To distinguish
between scattering states and resonances, it is necessary to
include multi-hadron operators in the matrix of correlation
functions.  However, we use only three quark operators in this
work.

\begin{table}[h]
\caption{The number of occurrences of each double valued irrep
$\Lambda=\{G_1,H,G_2\}$ of the octahedral group for different
values of continuum $J$ up to $\frac{11}{2}$.}
\begin{center}
\begin{tabular}{|c c c c c c c c|}
\hline $\Lambda$ & $J=$ & $\frac{1}{2}$ & $\frac{3}{2}$ &
$\frac{5}{2}$ & $\frac{7}{2}$ & $\frac{9}{2}$& $\frac{11}{2}$ \\
\hline
$G_{1}$ && $1$ & $0$ & $0$ & $1$ & $1$ & $1$ \\
 $H$     &&$0$ & $1$ & $1$ & $1$ & $2$ & $2$ \\
 $G_2$   &&$0$ & $0$ & $1$ & $1$ & $0$ & $1$ \\
\hline
 \end{tabular}
 \end{center}
 \label{tab:irreps}
 \end{table}
\section{Nucleon operators}
We construct baryon operators which transform as irreps of the
octahedral group as described in \cite{basak094}.  The basic
building blocks are covariantly displaced smeared quark fields
$(\tilde D^{(p)}_j\tilde \psi(x))_{Aa\alpha}$ with flavor $A$,
color $a$, and spin $\alpha$.  The smeared quark field $\tilde
\psi$ is displaced $p$ links in the $j$ direction
($j=0,\pm1,\pm2,\pm3$). Both quark smearing and gauge link
smearing are necessary to reduce noise and coupling to high energy
states \cite{lichtl_pos}.  We use Gaussian quark smearing and
stout link smearing \cite{morningstar}. From the displaced single
quark operators, we construct elemental operators
\begin{equation}
\Phi^{ABC}_{\alpha\beta\gamma;ijk}(x) = \epsilon_{abc}(\tilde
D^{(p)}_i\tilde \psi(x))_{Aa\alpha}(\tilde D^{(p)}_j\tilde
\psi(x))_{Bb\beta}(\tilde D^{(p)}_k\tilde \psi(x))_{Cc\gamma},
\end{equation}
where $\epsilon_{abc}$ is the antisymmetric Levi-Civita symbol. To
construct nucleon operators we project to $I=\frac{1}{2}, I_3=\frac{1}{2}$:
\begin{equation}
N_{\alpha\beta\gamma;ijk}=\Phi^{uud}_{\alpha\beta\gamma;ijk}-\Phi^{duu}_{\alpha\beta\gamma;ijk}.
\end{equation}

We use several different patterns of displacements for our three
quark operators, as summarized in Table \ref{tab:opforms}.  Taking linear combinations of
these elemental operators, we project to the irreps of the
octahedral group. This results in hundreds of nucleon operators in
each channel, which are then ``pruned'' to manageable sets as
described in \cite{lichtl_thesis}.

\begin{table}
\caption[captab]{Patterns of quark displacements employed in
elemental operators.  The displacement indices indicate the
direction of the gauge-covariant displacement for each quark;
$i,j,k=\{0,\pm 1,\pm2,\pm3\}$.  The quarks all have the same
displacement length.} \label{tab:opforms}
\begin{center}
\begin{tabular}{|cl|}
\hline
 Operator type &  Displacement indices\\
\raisebox{0mm}{\setlength{\unitlength}{1mm} \thicklines
\begin{picture}(16,10)
\put(8,6.5){\circle{6}} \put(7,6){\circle*{2}}
\put(9,6){\circle*{2}} \put(8,8){\circle*{2}}
\put(0,0){single-site}
\end{picture}}  & \raisebox{3mm}{$i=j=k=0$ }\\
\raisebox{0mm}{\setlength{\unitlength}{1mm} \thicklines
\begin{picture}(23,10)
\put(7,6.2){\circle{5}} \put(7,5){\circle*{2}}
\put(7,7.3){\circle*{2}} \put(14,6){\circle*{2}}
\put(9.5,6){\line(1,0){4}} \put(0,0){singly-displaced}
\end{picture}}  & \raisebox{3mm}{$i=j=0,\ k\neq 0$} \\
\raisebox{0mm}{\setlength{\unitlength}{1mm} \thicklines
\begin{picture}(26,8)
\put(12,5){\circle{3}} \put(12,5){\circle*{2}}
\put(6,5){\circle*{2}} \put(18,5){\circle*{2}}
\put(6,5){\line(1,0){4.2}} \put(18,5){\line(-1,0){4.2}}
\put(-1,0){doubly-displaced-I}
\end{picture}}  & \raisebox{2mm}{$i=0,\ j=-k,\ k\neq 0$} \\
\raisebox{0mm}{\setlength{\unitlength}{1mm} \thicklines
\begin{picture}(20,13)
\put(8,5){\circle{3}} \put(8,5){\circle*{2}}
\put(8,11){\circle*{2}} \put(14,5){\circle*{2}}
\put(14,5){\line(-1,0){4.2}} \put(8,11){\line(0,-1){4.2}}
\put(-5,0){doubly-displaced-L}
\end{picture}}   & \raisebox{4mm}{$i=0,\ \vert j\vert\neq \vert k\vert,
  \ jk\neq 0$}\\
\raisebox{0mm}{\setlength{\unitlength}{1mm} \thicklines
\begin{picture}(20,12)
\put(10,10){\circle{2}} \put(4,10){\circle*{2}}
\put(16,10){\circle*{2}} \put(10,4){\circle*{2}}
\put(4,10){\line(1,0){5}} \put(16,10){\line(-1,0){5}}
\put(10,4){\line(0,1){5}} \put(-5,0){triply-displaced-T}
\end{picture}}   & \raisebox{4mm}{$i=-j,\ \vert j\vert \neq\vert k\vert,
 \ jk\neq 0$} \\
\raisebox{0mm}{\setlength{\unitlength}{1mm} \thicklines
\begin{picture}(20,12)
\put(10,10){\circle{2}} \put(6,6){\circle*{2}}
\put(16,10){\circle*{2}} \put(10,4){\circle*{2}}
\put(6,6){\line(1,1){3.6}} \put(16,10){\line(-1,0){5}}
\put(10,4){\line(0,1){5}} \put(-5,0){triply-displaced-O}
\end{picture}}   & \raisebox{4mm}{$\vert i\vert \neq \vert j\vert \neq
  \vert k\vert,\ ijk\neq 0$}\\
  \hline
\end{tabular}
\end{center}
\end{table}

  To optimize the variational method,
we would like a set of low noise, linearly independent operators.
We first eliminate noisy operators based on the signal to noise
ratio in diagonal elements of the correlation matrix.  To test for linear independence of
a set of operators, we use the condition number of the normalized
correlation matrix on time slice $t=a_t$:
\begin{equation}
\hat C_{ij} = \frac{C_{ij}(a_t)}{\sqrt{C_{ii}(a_t)C_{jj}(a_t)}}.
\end{equation}
For completely orthogonal operators, the condition number is $1$
while for completely degenerate operators, the condition number
diverges.  We therefore seek a set of operators that minimizes
the condition number.  Pruning first within each operator type and
then across all remaining operators, we produce a set of $16$
optimal operators in each channel.  The operators used in this work are the same
as those in \cite{lichtl_thesis}.

We can write the correlator of two nucleon operators in terms of
three quark propagators
\begin{eqnarray}
\tilde G^{(ABC)}_{(\alpha |\bar\alpha )(\beta |\bar\beta )(\gamma
|\bar\gamma )} = &\sum_{\mathbf{x}}&\epsilon_{abc}\epsilon_{\bar a
\bar b \bar c}\tilde Q^{(A)}_{a\alpha|\bar a\bar\alpha}(\mathbf{x}
,t|\mathbf{x_0},0)
\times Q^{(B)}_{b\beta|\bar b\bar\beta}(\mathbf{x} ,t|\mathbf{x_0},0) \nonumber \\
&\times& Q^{(C)}_{c\gamma|\bar c\bar\gamma}(\mathbf{x}
,t|\mathbf{x_0},0),
\end{eqnarray}
where  $Q^{(A)}_{a\alpha|\bar a\bar\alpha}(\mathbf{x}
,t|\mathbf{x_0},0)$ is a single quark propagator of flavor $A$
from source site $\mathbf{x_0}$ at time $t=0$ to sink site
$\mathbf{x}$ at time $t$.  The nucleon correlation matrix is
\begin{eqnarray}
C_{ij} &=& c^i_{\alpha\beta\gamma}\bar
c^j_{\bar\alpha\bar\beta\bar\gamma}\Big\{\tilde
G^{(uud)}_{(\alpha|\bar\alpha)(\beta|\bar\beta)(\gamma|\bar\gamma)}
+G^{(uud)}_{(\alpha|\bar\beta)(\beta|\bar\alpha)(\gamma|\bar\gamma)}
-G^{(uud)}_{(\alpha|\bar\gamma)(\beta|\bar\beta)(\gamma|\bar\alpha)}
-G^{(uud)}_{(\alpha|\bar\beta)(\beta|\bar\gamma)(\gamma|\bar\alpha)}\nonumber
\\
&&-G^{(uud)}_{(\beta|\bar\beta)(\gamma|\bar\alpha)(\alpha|\bar\gamma)}
-G^{(uud)}_{(\beta|\bar\alpha)(\gamma|\bar\beta)(\alpha|\bar\gamma)}
+G^{(uud)}_{(\gamma|\bar\gamma)(\beta|\bar\beta)(\alpha|\bar\alpha)}
+G^{(uud)}_{(\gamma|\bar\beta)(\beta|\bar\gamma)(\alpha|\bar\alpha)}
\Big\},
\end{eqnarray}
where the $c^i_{\alpha\beta\gamma}$ are the coefficients to project
the operators to irreps of the octahedral group.  The ``$i$''
superscript identifies each individual nucleon operator used to
construct the matrix.

\section{Computational method}
\subsection{Variational Method}
 We use the variational method \cite{luscher,lichtl_thesis} to extract the excited spectrum from the matrix of correlation functions, numerically
  solving the generalized eigenvalue problem
\begin{equation}
 C_{ij}^{(\Lambda)}(t)v_{j}^{(n)}(t,t_0) = \alpha^{(\Lambda)}_n(t,t_0)C_{ij}^{(\Lambda)}(t_0) v_{j}^{(n)}(t,t_0),
\end{equation}
where $n$ labels the eigenstates and $\Lambda$ the double valued
irrep of the octahedral group. To reduce instabilities in the
eigenvectors due to degeneracies and numerical uncertainties, we
solve the eigenvalue problem on a single time slice $t^*$. This
time is selected to be as small as possible to minimize the noise,
but large enough that the eigenvectors have stabilized.  The
correlator on all other time slices is then rotated to this fixed basis
of eigenvectors.

The diagonal elements of the rotated correlation matrix $\tilde C_{ij}(t)$ are
related to the energies by
\begin{equation}
 \tilde C_{ii}^{(\Lambda)}(t)\simeq e^{-E_i(t-t_0)}\left(1+\mathcal{O}\left(e^{-|\delta E|t}\right)\right)
  +\sum_{n\neq i} \alpha_n(t) e^{-E_n(t-t_0)},
\end{equation}
where $\delta E$ is the difference between $E_i$ and the next
closest energy.  The summation is a correction which results from
the fact that we diagonalize only at $t^*$.  On other time slices
we expect a small overlap between the basis vectors and all other
energy eigenstates. However, the coefficients $\alpha_n(t)$ should
be negligible near $t^*$. We can calculate the fixed eigenvector
effective energy via
\begin{equation}
E_i^{eff}(t) = \ln \left[ \frac{\tilde
C_{ii}^{(\Lambda)}(t)}{\tilde C_{ii}^{(\Lambda)}(t+1)}\right].
\end{equation}

\subsection{Lattice Action}
We used $24^3\times 64$ anisotropic lattices with the temporal
lattice spacing $a_t$ three times finer than the spatial lattice
spacing.  Gauge configurations were generated using the Wilson
action. The scale was set with
the Sommer parameter.  We analyzed an ensemble of $430$
configurations with $M_\pi=400$ MeV.

\subsection{Fit Details}
Because baryon operators create a baryon and annihilate an
antibaryon and we use antiperiodic temporal boundary conditions,
if we create a baryon state at $t=0$, we also create a state which
propagates backwards in time from the opposite temporal end of the
lattice. Because fermions and antifermions have opposite intrinsic
parity, the antibaryon propagating backward in time has parity
opposite to that of the  fermion propagating forward in time.

     Since the signals decay exponentially,
the backward-propagating signal
    will only be above the noise level for time slices above some threshold value of time.
If the forward signal has an effective energy plateau in a region
below this threshold then we can extract the energy of the state by
fitting the correlation
  function  with an exponential decay through the plateau without any interference from the backward propagating signal.
    This was, in fact, the case for all channels except for $G_{1u}$.
      In this channel, the backward propagating signal was dominated by the $G_{1g}$ ground state,
       the lowest energy state in the spectrum.  For our lattice, the backward-propagating
       signal decayed slowly enough and the
       temporal extent was small enough (due to the anisotropy) that the $G_{1u}$ effective energies had significant backward contamination.
To extract the energy levels, we fit the correlation
functions as a forward exponential decay plus a backward
exponential decay  which was constrained by the $G_{1g}$ ground
state.

We performed fully correlated $\chi^2$ minimization fits. For each
state a fit range was selected such that the $\chi^2$ was
minimized, the quality factor $Q$ was maximized, the corresponding
effective energy plot plateaued in the fit range, and the fit
parameters were stable under small variations in the fit range. As
stated above, each channel except for the $G_{1u}$   was modeled as
\begin{equation}
 \tilde C^{(\Lambda)}_{ii}(t) = A e^{-E^{(\Lambda)}_i (t-t_0)}.
\end{equation}
The $G_{1u}$ correlation functions were fit simultaneously with the
$G_{1g}$ ground state, constraining the $G_{1g}$ ground state
energy to be equal to the energy of the backward-propagating $G_{1u}$
state:
\begin{eqnarray}
 \tilde C^{(G_{1u})}_{ii} &=& A e^{-E^{(G_{1u})}_i(t-t_0)} + B e^{-E^{(G_{1g})}_0 (T-t)} \nonumber\\
 \tilde C^{(G_{1g})}_{00} &=& D e^{-E^{(G_{1g})}_0 (t-t_0)}
\end{eqnarray}
We estimated the uncertainty in the energy through a jackknife
analysis.  We repeated the fit for each single elimination jackknife sample. The average
energy and its jackknife error are reported.

\subsection{Filtering}
The presence of the backward-propagating state in the $G_{1u}$
channel made the L\"uscher method, the diagonalization of the
correlation matrix on each time step, unworkable because it led to
large numerical instabilities in the eigenvectors.  We tested a
method based on filtering out the backward signal prior to
diagonalization.  In a time interval where the backward signal is
simply the ground state of the opposite parity channel we model the unrotated correlation matrix as
\begin{equation}
 C_{ij}^{(\Lambda)}(t)=\sum_{n} A_n e^{-E^\Lambda_n(t-t_0)} + B e^{-E_0^{\Lambda_c}(T-t_0)}.
\end{equation}
We define the filtered correlator as
\begin{eqnarray}
 C_{filt,ij}^{(\Lambda)}(t,t_1) &=& C_{ij}^{(\Lambda)}(t) - C_{ij}^{(\Lambda)}(t_1) +(1-e^{-E^{\Lambda_c}}_0) \sum_{j=t+1}^{t_1} C_{ij}^{(\Lambda)}(j) \\
&=& \sum_n
A_n\left[1+\frac{1-e^{-E^{\Lambda_c}_0}}{e^{E^\Lambda_n}-1}\right]\left(e^{-E^{\Lambda}_n(t-t_0)}-
e^{-E^{\Lambda}_n(t_1-t_0)}\right),
\end{eqnarray}
where $t_1$ is a time where the backward signal is, in fact,
described by a single exponential. The filtered correlators consist
of the forward signal plus a constant term. The diagonalization of
the filtered correlators using the L\"uscher method produced
stable eigenvectors, and the energy of the states could be
extracted by fitting the principal correlation functions to a
single exponential decay with a constant term.  This method did
not produce any significant improvement over the results from the
fixed eigenvector method, but we point out that the filtering is
necessary to extract the excited spectrum using the L\"uscher
method when there is significant backwards contamination in the effective energy.
\section{Results}
\begin{figure}
\begin{center}
\includegraphics[width=0.6\textwidth,clip=true]{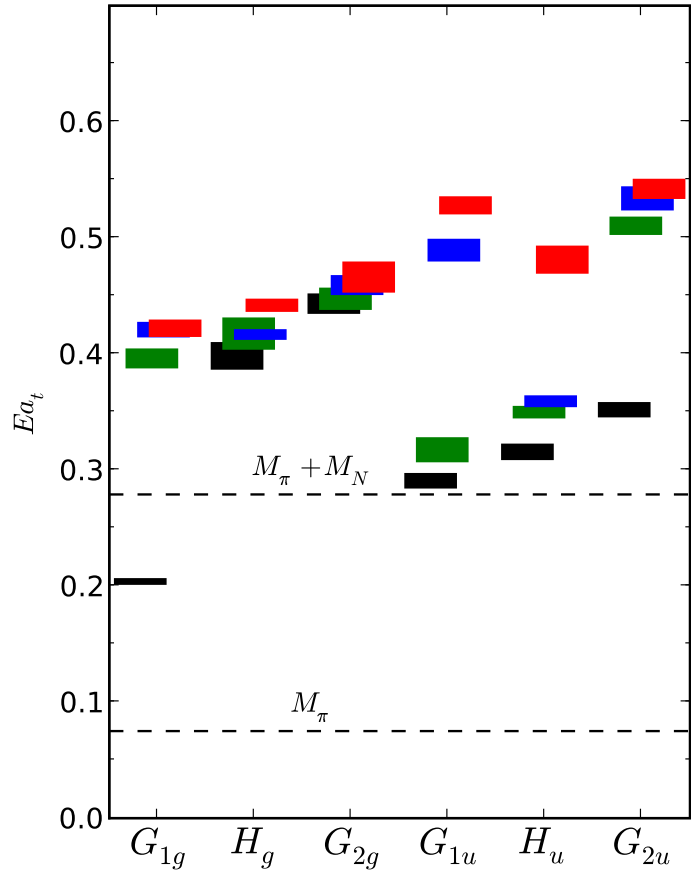} 
\end{center}
\caption{The nucleon spectrum obtained for each symmetry channel for
$24^3\times64$ $N_f=2$ lattice QCD data at $m_\pi = 400$ MeV.  Errors are indicated by the vertical
size of the box.} \label{fig:boxplot_m400}
\end{figure}

Four states were obtained in each channel.  A plot of the spectrum is shown in
Fig. \ref{fig:boxplot_m400}.  In the positive parity channels, we can identify the $G_{1g}$ ground state with the nucleon.  Due to the high degree of degeneracy in the excited positive parity channel, it is difficult to identify the excited states.

   Results for the low energy negative-parity excited states allow for some interesting interpretations.
  In the $G_{1u}$ channel, we see the energy of the lowest energy states are close to the sum of the pion and nucleon mass.  This raises the possibility that one of these states is, in fact, a pion-nucleon state.  Even if this is the case, the lowest negative-parity nucleon state in $G_{1u}$ is below the first excited positive-parity state in $G_{1g}$ . This does not match the physical spectrum where the lowest negative parity resonance, the $N^*(1535)$  is above the first excited positive parity state, the $N(1440)$.

  In the $H_u$ channel, we expect the lowest state to correspond to a spin $\frac{3}{2}^-$ state.  We interpret the lowest $H_u$ state as the $N(1520)$.  In the $G_{2u}$ channel we see that the lowest-energy is degenerate with a partner $H_u$ state, with no state in the $G_{1u}$ channel at the same energy. This is the signature of a spin $\frac{5}{2}^-$ state.  Two of the six components needed for a spin $\frac{5}{2}^-$ state are occur in  the $G_{2u}$ channel and four in the $H_u$ channel.  The $G_{2u}$ and $H_u$ states corresponding to the $\frac{5}{2}^-$ must be degenerate in the continuum limit. Moreover, there should not be a $G_{1u}$ state that is degenerate with these two states as that would indicate the possible presence of a spin $\frac{7}{2}^-$ state. These conditions are satisfied near $E a_t = 0.35$ giving clear evidence for a $\frac{5}{2}^-$ state.  We interpret this to be the lowest $\frac{5}{2}^-$ state in the physical spectrum, the $N(1675)$.

    A $\frac{5}{2}^-$ state has not been seen in any earlier work. Our quenched QCD analysis had three degenerate states (within errors): $G_{2u}$ , $H_u$ and $G_{1u}$. That pattern had two possible interpretations: a single spin $\frac{7}{2}^-$ state or an accidental degeneracy of a spin $\frac{5}{2}^-$ state and a spin $\frac{1}{2}^-$ state.

\section{Acknowledgments}
This work was done using the Chroma software
suite~\cite{Edwards:2004sx} on clusters at Jefferson Laboratory using
time awarded under the USQCD Initiative. This research used resources
of the National Center for Computational Sciences at Oak Ridge
National Laboratory, which is supported by the Office of Science of
the U.S. Department of Energy (DOE) Under contract DE-AC05-00OR22725
under the US DOE INCITE 2007 Program. This research was supported in
part by the National Science Foundation through Teragrid Resources
provided by the San Diego Supercomputing Center (Blue Gene).  This work was also partially supported by National Science Foundation award PHY-0653315 and by the Department of Energy under contracts DE-AC05-06OR23177 and DE-FG02-93ER-40762.  The U.S. Government retains a non-exclusive,
paid-up, irrevocable, world-wide license to publish or reproduce this
manuscript for U.S. Government purposes.

\end{document}